# QCD with dynamical Wilson fermions— first results from SESAM*

SESAM-Collaboration:
U. Glässner[b], S. Güsken[b], H. Hoeber[a], Th. Lippert[a], X. Luo[a], G. Ritzenhöfer[a], K. Schilling[a,b], and G. Siegert[a].

[a]HLRZ c/o Forschungszentrum Jülich, D-52425 Jülich, and DESY, D-22603 Hamburg, Germany,

[b]Physics Department, University of Wuppertal, D-42097 Wuppertal, Germany.

First results of a recently started simulation of full QCD with two flavours of sea-quarks at a coupling of $\beta = 5.6$ on a $16^3 \times 32$ lattice are presented. Emphasis is laid on the statistical significance that can be achieved by an "integrated luminosity" of 140 TFlop×hrs, for Hybrid Monte Carlo simulations at four intermediate values of $\frac{m_\pi}{m_\rho}$. The simulation takes place on the Quadrics QH2 at DESY/Zeuthen and DFG/Bielefeld. The performance is optimized by means of BiCGStab and the chronological inversion method of Brower et al. We discuss the systematic errors arising from lack of the molecular dynamic's reversibility on the 32-bit QH2. For plaquette and meson correlators we find integrated autocorrelation times of $< 20$ units of molecular dynamics time and exponential autocorrelation times of about 50 units. Using these results we perform preliminary measurements of the central potential and $\pi$ and $\rho$ correlators on independent configurations and obtain first estimates of the lattice spacings at three values of the dynamical hopping parameter.

## 1. INTRODUCTION

There is a growing need for progress in the treatment of full QCD with Wilson fermions at zero temperature as quenched simulations are becoming more and more precise and we are approaching the point where the effects of dynamical fermions are expected to become visible.

It is well known from the pioneering large scale computer experiments on Connection Machines[1,2], however, that dynamical Wilson fermions are particularly difficult to simulate, and it remained unsettled at the time whether the standard Hybrid Monte Carlo (HMC) algorithm is suited at all to provide us with a sufficient sample of decorrelated configurations at sufficiently weak coupling[3]. With the advent of powerful dedicated computers like APE100, this issue has been readdressed, and an autocorrelation time in the HMC series has been quoted (from a sample of 500 trajectories) to be of order 50[4].

Nevertheless, for a physics analysis in full QCD, one still lacks a reliable determination of the autocorrelation time $\tau$ and a sufficiently large sample of decorrelated field configurations. Such a program requires for sure the generation of $\mathcal{O}(100 \times \tau)$ trajectories. To be specific: one would need envisage a computer experiment with order 20.000 trajectories, to reach significant results.

Apart from the physics motivation, such an experiment is highly desirable as reference point to assess the benefit from 'improved' algorithms which are currently being studied[5].

In this spirit we have formed a collaboration with the magic name 'SESAM' which is meant to unlock the door—through a statistically significant sample—to behold **S**ea Quark **E**ffects on **S**pectrum **A**nd **M**atrix Elements. Our magic key carries the name of QH2 which delivers cheap real Gigaflops, as installed by DESY in Zeuthen and DFG in Bielefeld.

## 2. EXPERIMENTAL PROCEDURES

We employ the standard Hybrid Monte Carlo algorithm[6,7] (HMC) with $n_f = 2$ Wilson fermions on a lattice of size $16^3 \times 32$, with fermionic boundary conditions anti-periodic in time. The results of the SCRI-group on the full-QCD $\beta$-function [3] suggest to go beyond $\beta = 5.5$

*Poster presented by H. Hoeber and talk presented by Th. Lippert

in order to stay away from the strong coupling regime. To remain with a reasonable physical lattice size, we chose to work at $\beta = 5.6$. With algorithmic improvements and a fast implementation of the HMC on the QH2 (to be described below), we achieve a sustained HMC-performance of nearly 8 Gflops at a QH2 peak-rate of 12.8 Gflops.

Given these parameters and the performance data for the inversion part of the HMC quoted in Refs. [1,2,8], we could estimate, during the layout stage of the computer experiment, the number of trajectories that would be achieved on QH2, given a years running time, as a function of $\frac{m_\pi}{m_\rho}$. We have normalized the trajectory length to $T = 1$, extrapolated linearly to a residue of $r = \frac{||M^\dagger Mx-\phi||}{||x||} = 10^{-8}$ (see section 4) and—using the scaling laws of Ref. [1]— further extrapolated to our lattice volume of $V = 16^3 \times 32$.

This results in the performance plot of Fig. 1. After 100 days of running time, it is gratifying to find ourselves (SESAM) in accord with these expectations.

In the tuning of $\kappa$ towards its chiral regime, we should aim at small values of $\frac{m_\pi}{m_\rho}$, while ensuring finite-size effects to remain tolerable, to say $m_\pi L \geq 4$. This requirement endows the available $\frac{m_\pi}{m_\rho}$ range with a lower limit which marks the 'chiral target region'. In Fig. 1 the latter is indicated by the vertical lines.

During the tuning phase, we approached the 'target region' in an iterative procedure. Eventually we decided to generate a mass trajectory of 4 dynamical $\kappa$-values, $\kappa = 0.1560, 0.1565, 0.1570, 0.1575$. In the first hundred days of dedicated run-time, we have produced around 5000 trajectories at three different values of the mass of the sea-quark, see Table 1.

## 3. ACCELERATING HMC

Lattice QCD can profit considerably from the development of new types of iterative solvers like BiCGStab[9] or QMR[10] as has recently been shown in Refs. [11–14].

In the context of HMC, we apply BiCGStab in a two step procedure as is the case with MR.

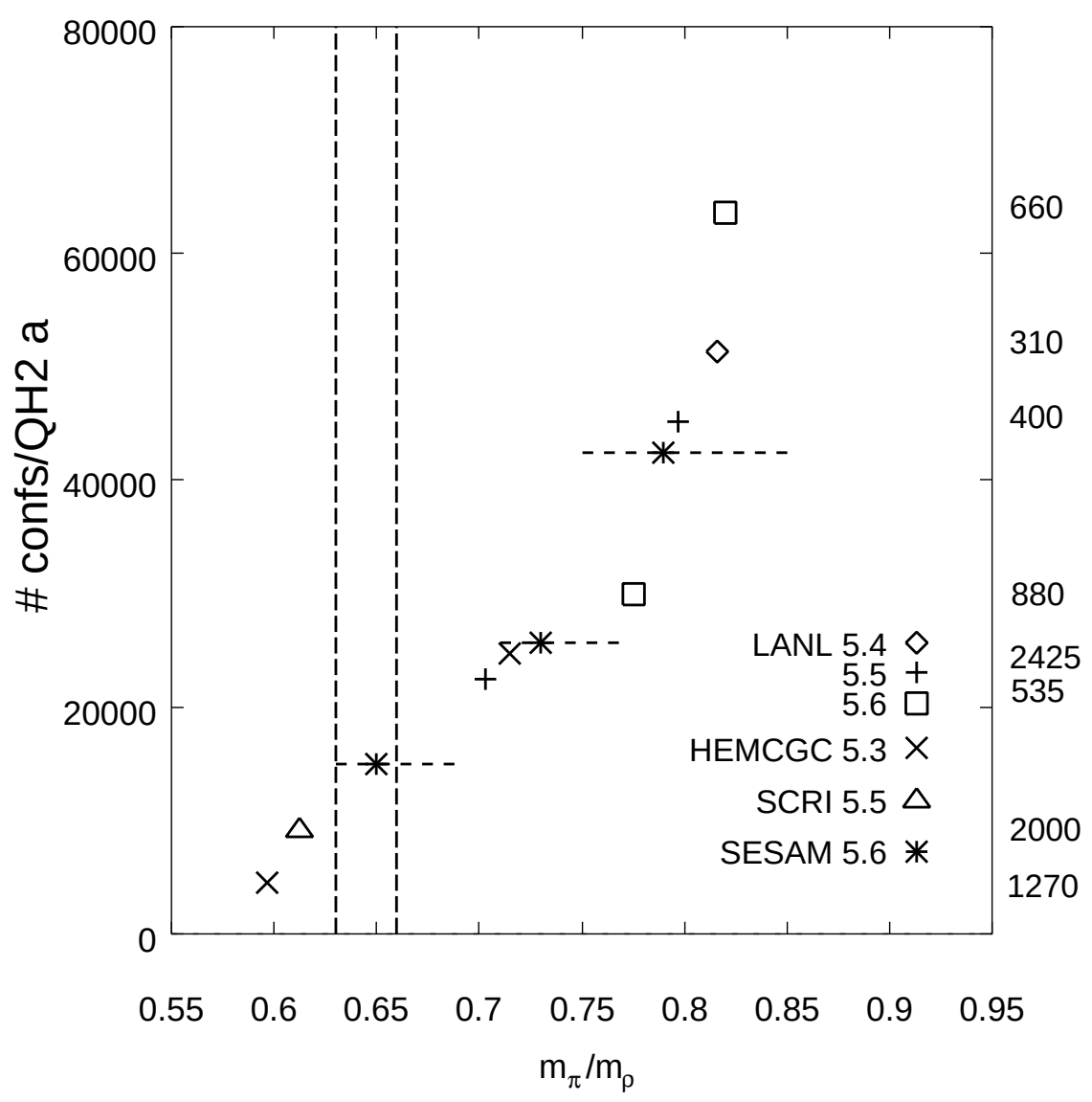

Figure 1. Performance plot: number of configurations achievable in one year of QH2 run-time, plotted against $\frac{m_\pi}{m_\rho}$, as estimated from preceding computer experiments. The plot also shows, on the right hand scale, the actual numbers of trajectories achieved by Refs. [1,2,8].

Repeatedly, during the molecular dynamics evolution, the equation $M_e^\dagger M_e x = \phi$ has to be solved, with $M_e$ being the even-odd preconditioned Wilson fermion matrix $M_e = 1 - \kappa^2 D_{eo} D_{oe}$, where the Wilson fermion matrix is given as

$$M = \begin{pmatrix} \mathbf{1} & -\kappa D_{oe} \\ -\kappa D_{eo} & \mathbf{1} \end{pmatrix}. \quad (1)$$

The solution proceeds in two-step mode:

$$\text{first:} \quad M_e^\dagger y = \phi, \quad \text{second:} \quad M_e x = y, \quad (2)$$

where both equations are solved with equal accuracy, see below. It has been shown in Ref. [11] that the resulting solution vectors $x$ are equal within round-off errors compared to a direct CG solution of $M_e^\dagger M_e x = \phi$.

The multiplication by the matrix $M_e$ requires 28 milli-seconds on the QH2 for the $16^3 \times 32$ lattice. At $\kappa = 0.1575$ we need about 40 minutes to run one trajectory of 100 molecular dynamics steps.

Fig. 2 plots the accumulated residual $r$ as function of the cpu-time, computed on the QH2 for a typical thermalized configuration at our smallest bare quark mass. We found that overrelaxed MR outperforms CG, yet BiCGStab beats ORMR. This is shown in the figure and is accordance with Ref. [11]. The gain is typically 30 % compared to the performance of BiCGStab. We note that the true residual $r$ runs into a plateau at $r = 10^{-6}$ as the matrix multiplication is carried out in single precision. Note that global operations inside the inverter are emulated in Kahan precision[15], however.

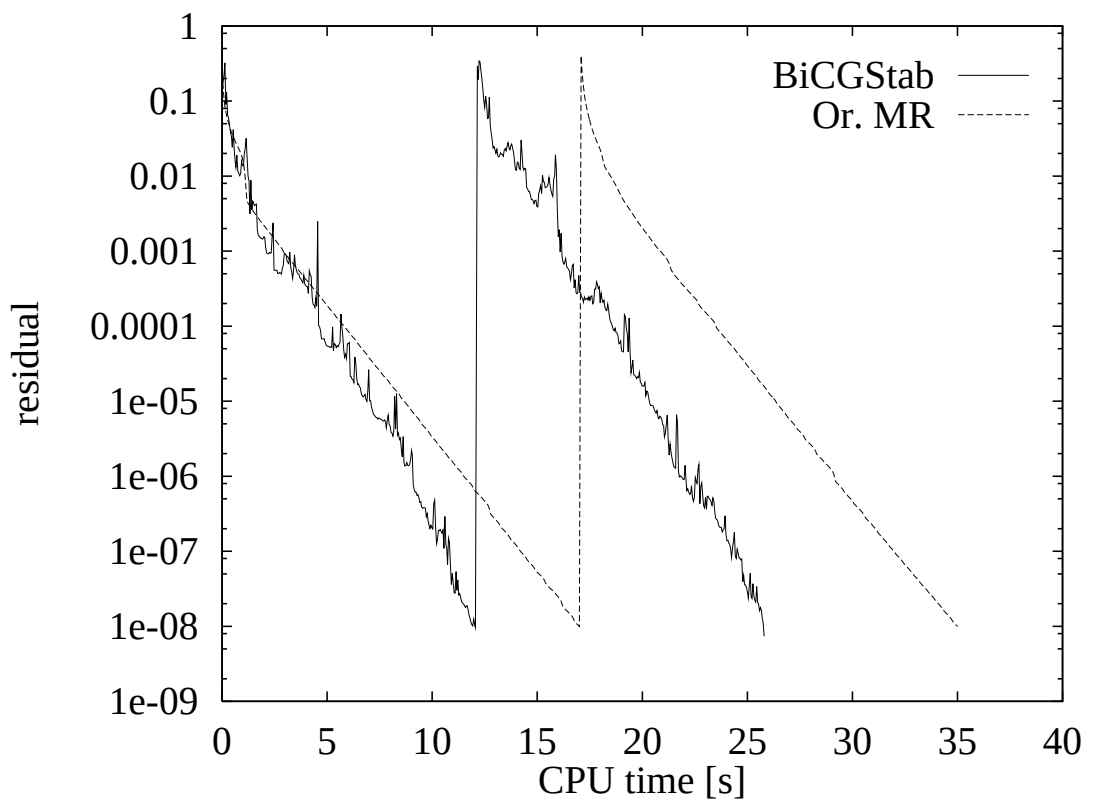

Figure 2. Accumulated residual $r$ as function of the cpu-time in the two step inversion of $M_e^\dagger M_e x = \phi$.

A method to further improve any inverter's performance has been presented in Ref. [16]: assuming a smooth behaviour in the molecular dynamics evolution of the solution vector $x$ for small molecular dynamics time steps $dt$, one can guess the vector $x_{n+1}$ at time step $n+1$ from the preceeding vectors $x_n, x_{n-1} \dots x_1$. We use the simple polynomial formula as given in Ref. [16]:

$$x_{n+1}^{\text{trial}} = \sum_{i=1}^{n} c_i \, x_i , \tag{3}$$

where

$$c_i = (-1)^{i-1} \begin{pmatrix} n \\ i \end{pmatrix} . \tag{4}$$

In second order, this procedure had been suggested before in Ref. [7].

In Fig. 3, we plot the number of iterations along the molecular dynamics evolution. We have em-ployed an extrapolation up to 8th order to determine the starting vector. Again plotting a typical situation, we see that nearly a factor of 2 can be gained in the number of iterations after the start-up phase. We note that there are a few particular trajectories that cannot be accelerated using this method. For $\kappa = 0.1575$, we go to 11th order.

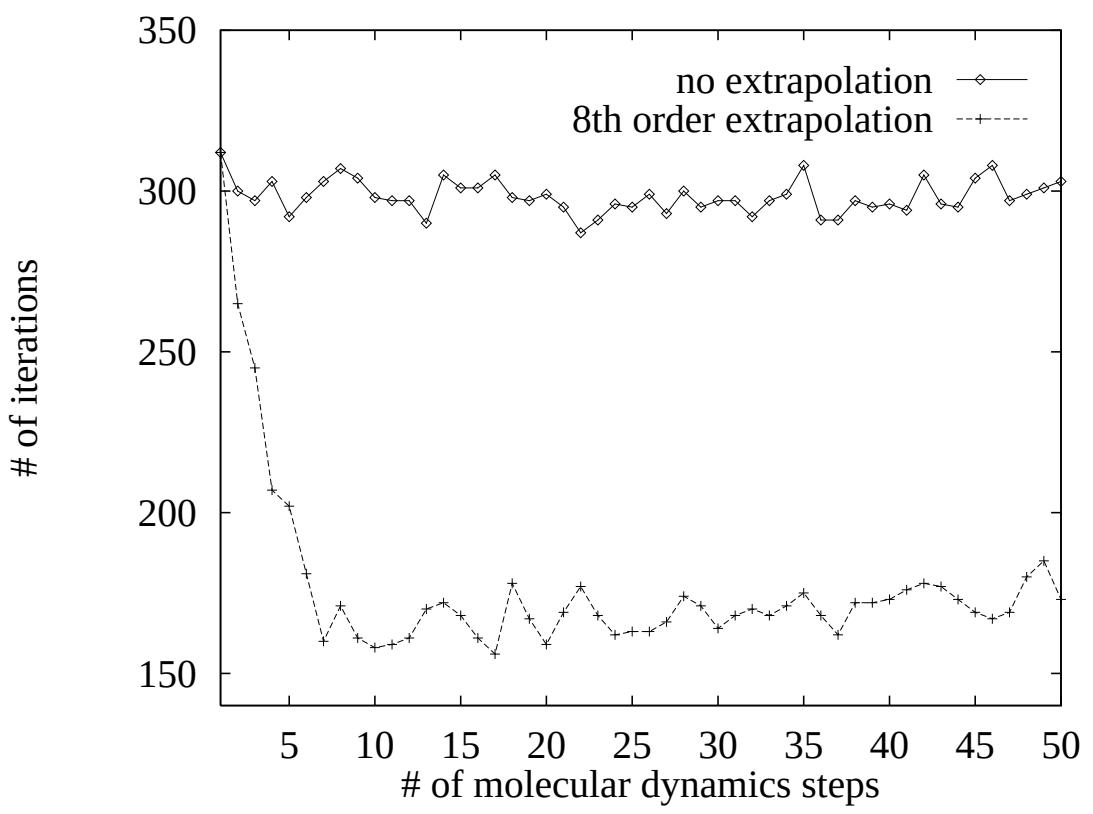

Figure 3. Extrapolation of the start vector $x$ to 8th order in $dt$ at a $\kappa$-value of 0.1570.

## 4. REVERSIBILITY

It has been pointed out recently[17] that HMC carries a positive Lyapunov exponent. Therefore one has to worry that the ensueing irreversibility does not spoil detailed balance in the HMC process on the 32-bit Quadrics machine. It is important to make sure that the Lyapunov amplitude remains small along a trajectory.

A measure for this is the norm of the mismatch[17], between an initial and a final gauge configuration $\{U_i\}$ and $\{U_f\}$ 'around a closed loop':

$$||dU|| = \sqrt{\frac{1}{4V}\sum_{x,\mu}[U_{\mu,f}(x) - U_{\mu,i}(x)]^2}. \qquad (5)$$

$\{U_f\}$ is the result of a molecular dynamics integration, where, after $N_{md}$ steps (turning-point), the momenta and the time direction are reversed, $p \to -p, dt \to -dt$, such that after another $N_{md}$ steps we should arrive at $U_f = U_i$—given the numerical procedure to be exact. Numerical errors, however, can either be amplified or suppressed by the dynamics and—if the asymptotics of $||dU||$ is defined as

$$||dU|| \propto e^{\nu\, dt\, N_{md}} \qquad (6)$$

—either a positive of negative exponent $\nu$ will be found.

In Fig. 4, we have drawn the norm $||dU||$ as function of the number of molecular dynamics steps $N_{md}$ at the turning-point. On a thermalized $16^3 \times 32$ lattice for both quenched and full QCD, we find an equal value of $\nu = 0.75$.

With regard to the global HMC Monte Carlo decision, it is important to study the impact of the above mismatch on $\Delta S$. The violation of detailed balance is expressed by the factor $Z$,

$$W(U)\, P(U \to U') = W(U')\, P(U' \to U) \cdot Z, \qquad (7)$$

in terms of the fixed point distribution $W(U)$ and the evolution probability $P$, where

$$Z = \exp(\delta(\Delta S)). \qquad (8)$$

$\delta(\Delta S)$ is the (systematic) error in $\Delta S$ and a monitor for the violation of detailed balance.

In order to acquire a feeling for the relevance of this error and its dependency on the precision of the inversion on the QH2, we computed $\Delta S$ along a closed trajectory loop, integrating forward and then backwards in time, with $N_{md} = 100$ for $10^{-4} \geq r \geq 10^{-8}$ on one thermalized $16^3 \times 32$ lattice. During this operation, the global action is carefully tracked in double precision arithmetic[15]. The resulting mismatch value in $\Delta S$ that we denote by $\delta(\Delta S)$ is plotted *vs.* $r$ in Fig. 5. Note that this calculation is done with an 11th order guess as described above. While $\Delta S \simeq 1$, we find $\delta(\Delta S)$ to settle on the $10^{-2}$ level at $r < 10^{-7}$, the saturation being indicative for the impact of the 32 bit accuracy to that very precision in $S$. As the quantity $\delta(\Delta S)$ is a measure for the size of the

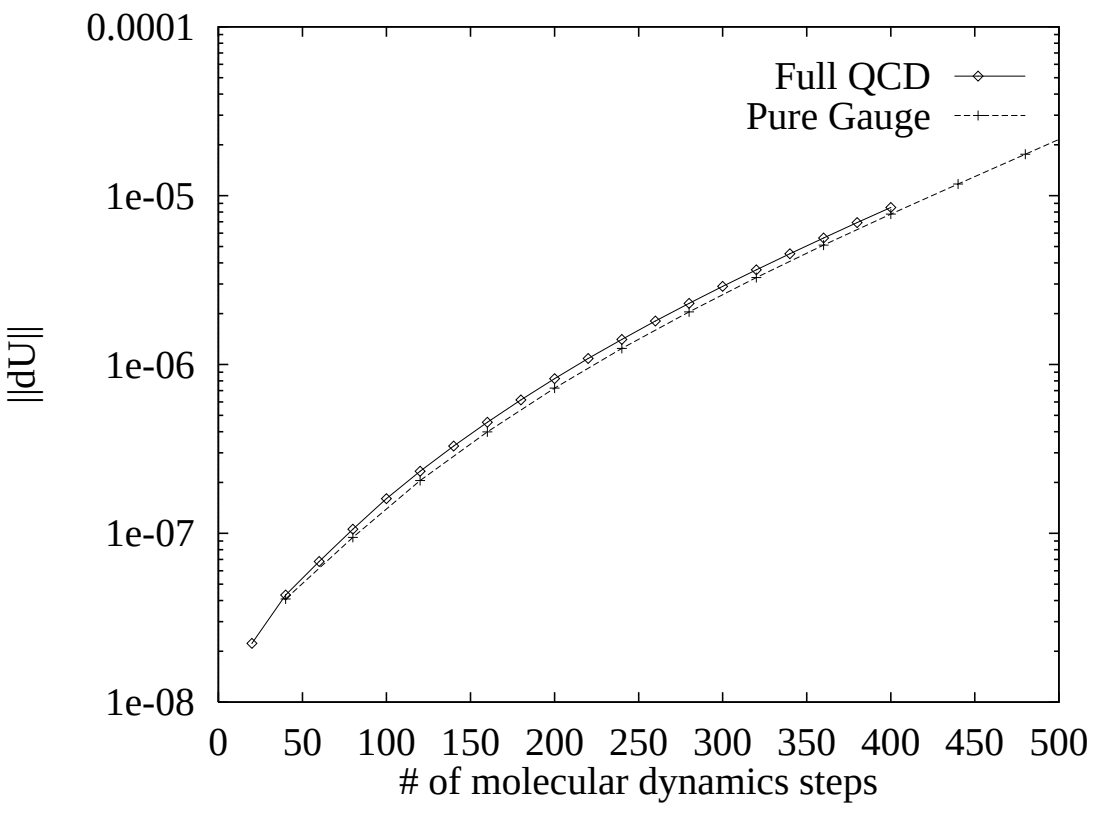

Figure 4. $||dU||$ as function of $N_{md}$ at the turning-point, $r = 10^{-8}$.

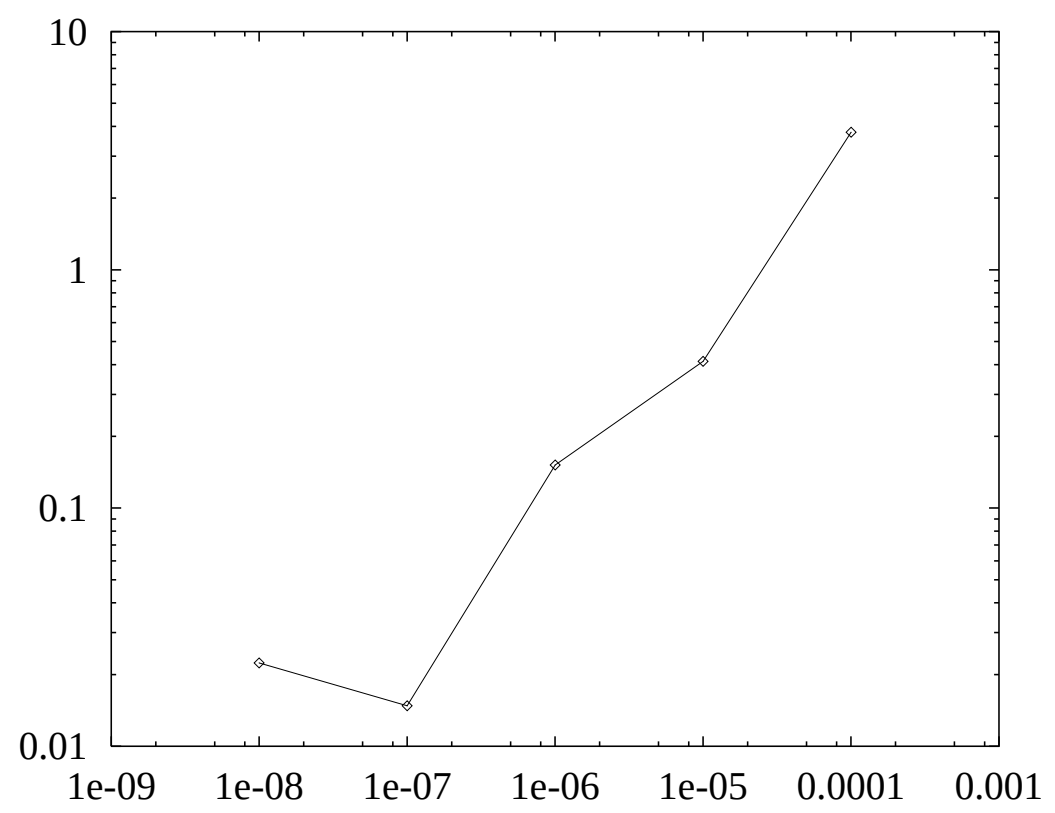

Figure 5. $\delta(\Delta S)$ as function of $r$.

Lyapunov amplitude we conclude that the dangerous instability appears to be rather *marginal* under the conditions of our HMC process, with its stochastic pivot amplitude $\Delta S$! Note that during the actual simulation we used $r = 10^{-8}$.

Alternatively, one might argue that the size of the expectation value of $\exp(-\Delta S) - 1$ is a measure for the effective violation of reversibility along the time history of an HMC run. We found $\langle \exp(-\Delta S) - 1 \rangle = 0.009 \pm 0.022$ on a contiguous sample of 1000 configurations for $\kappa = 0.1575$!

## 5. RESULTS

Table 1 lists the number of trajectories that have been generated in about 100 days cpu-time on the QH2. The second row gives the numbers of configurations that we consider to be thermalized.

Table 1
Number of trajectories generated so far. The last row shows the number of thermalized configurations with anti-periodic boundary conditions.

| $\kappa_{\rm sea}$ | total # of trajectories | # of thermalized trajectories |
|---|---|---|
| 0.156 | 564 | 160 |
| 0.1565 | 0 | 0 |
| 0.157 | 2478 | 1800 |
| 0.1575 | 1795 | 1400 |

In Table 2 we give the HMC run parameters that we chose for so far three $\kappa_{\rm sea}$ values at $\beta = 5.6$.

### 5.1. Autocorrelation

Knowledge of the autocorrelation is instrumental for assessing the statistical relevance of a given sample of trajectories. As we archive all configurations we are in the position to compute the autocorrelation of any operator. During production we have monitored *in situ* the plaquette and the pion propagator (on one time slice) for our two smallest quark masses.

Table 2
Run parameters of HMC simulation. In the second half of the runs we have switched over to variable $N_{md}$ uniformly distributed with a width of 20.

| $\kappa_{\rm sea}$ | $N_{md}$ | $dt$ | $r$ | acc/% | ⟨ # it.⟩ |
|---|---|---|---|---|---|
| 0.156 | 100 | 0.01 | $10^{-8}$ | 85 % | 100 |
| 0.157 | 100 | 0.01 | $10^{-8}$ | 82 % | 166 |
| 0.1575 | 100 | 0.01 | $10^{-8}$ | 76 % | 299 |

The exponential autocorrelation time, $\tau_{exp}$, is defined as the flattest slope in the logarithm of the autocorrelation function and is related to the length of the thermalization phase[18]. In principle, it is universal, i. e. independent on the particular underlying operator. In practice one cannot guarantee the 'slowest' mode to be really the slowest. In this paper we assume the thermalization to be $5 \times \tau_{exp}$, i. e. we require the influence of the initial configuration onto the "first" thermalized configuration to be of order $O(\exp(-5))$.

The statistical error of an operator $\mathcal{O}$ is characterized by the (operator dependent) integrated autocorrelation time. Fortunately, slow modes go along with small amplitudes, and therefore we can determine a reasonable estimate of $\tau_{int}$ from the behaviour of the leading modes only.

A sensible measurement of the autocorrelation function requires a sufficient length of the Markov chain. This fact prevented a reliable measurement of HMC generated configurations on large lattices up to now. In Fig. 6, we have plotted the autocorrelation function of the plaquette $P$. In order to demonstrate the impact of the sampling on the correlation measurement we display $C(t)$ vs. $t$ as a function of the sample size. We find that there exists a threshold in sample size (at about 1200 trajectories!) above which a slow mode can be identified in our time series analysis: from thereon the slope of $\log(C(t))$ is unravelled from the noise[2].

Table 3 presents the values for the autocore-lation times we computed so far. Our determinations of the exponential and integrated autocorrelation times are based on our thermalized samples at $\kappa = 0.1570$ and 0.1575 (see Table 1) for the plaquette $P$ and the pion propagator at $T = 16$. We performed uncorrelated fits to $\log(C(t))$ to extract $\tau_{exp}$ and estimated the integrated autocorrelation time $\tau_{int}$ according to $\tau_{int} = \lim_{t\to\infty} A(1 - \exp(-t/\tau))$.

[2]For the estimation of the error of $C(t)$ we took into account the integrated autocorrelation time of the autocorrelation function itself!

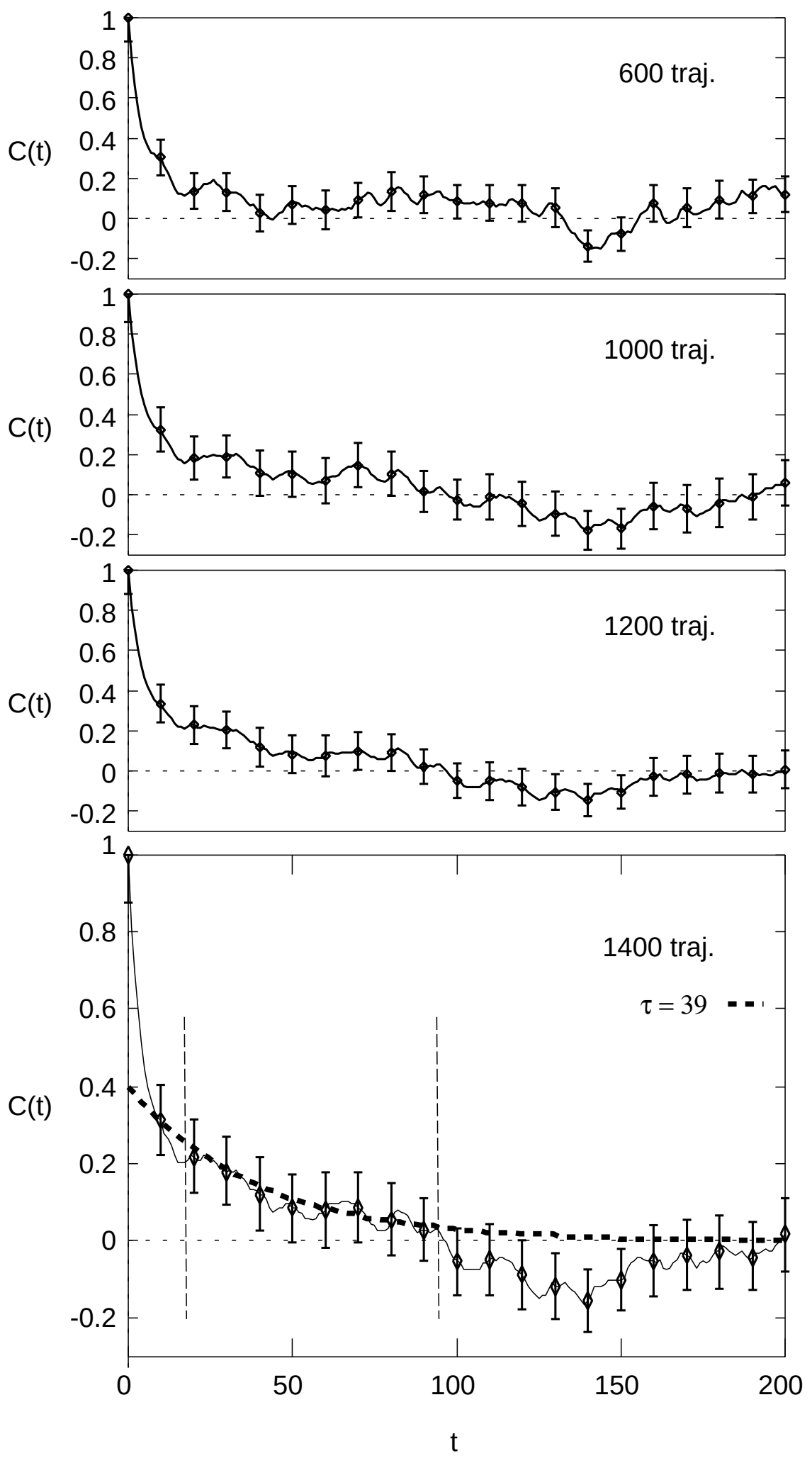

Figure 6. Autocorrelation function $C(t)$. The curves correspond to various lengths of the contiguous sample. A reliable determination of the slowest dynamical mode can be performed with a sample of 1400 trajectories, where $C(t)$ follows a single exponential in the range of 20 and 95 time units.

It is gratifying to find $\tau$ well below the value of 100 trajectories! On the basis of this result, it appears realistic, with the present means, to achieve a final sample of 100 independent configurations at each of the 4 $\kappa$ values quoted in Table 4.

Table 3
Exponential and integrated autocorrelation times of plaquette and pion at $\kappa = 0.1570$ and 0.1575.

| $\kappa_{\rm sea}$ | 0.1570 | 0.1575 |
|---|---|---|
| $\tau_{int}$ of plaquette | 12.3(5) | 14.8(8) |
| $\tau_{int}$ of pion propagator | 6.9(8) | 9.2(9) |
| $\tau_{exp}$ of plaquette | 31(17) | 39(31) |
| $\tau_{exp}$ of pion propagator | 34(15) | 57(26) |

Table 4
Estimate of the number of configurations of our mass trajectory that can be generated within 140 TFlop×hrs cpu time on the QH2 to end up with $\approx$ 100 independent configurations at each $\kappa$ value.

| $\kappa_{\rm sea}$ | total # of trajectories |
|---|---|
| 0.1560 | 4000 |
| 0.1565 | 5000 |
| 0.1570 | 6000 |
| 0.1575 | 8000 |

### 5.2. Static Potential

We[3] compute the static (spin independent) potential $V(R)$ from the path-ordered products of link variables around space-time rectangles performing measurements on configurations separated by 50 units of time - this corresponds to more than twice the integrated plaquette autocorrelation time (see section 5.1).

[3]This part of the work is done in collaboration with A. Wachter.

In order to enhance the ground-state signal, we use a local smearing procedure on the spatial links as depicted below.

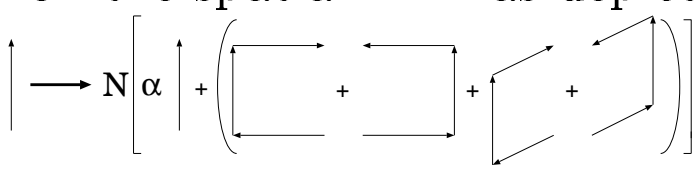

Here, N is a projection operator onto SU(3); the smearing parameter $\alpha$ is set to $\alpha = 2$. 100 smearing iterations are performed giving significant improvement of the signal for all distances. Overlaps are bigger than 90 % for most $R$.

Off-axis measurements are included for improved spatial resolution; this also allows an investigation of rotational invariance restoration. We perform measurements at 36 different values of $R$.

The potential is defined through local masses; for fixed time $T$ : $V^T(R) = \log \frac{\mathcal{W}(R,T)}{\mathcal{W}(R,T+1)}$. We find good plateaus for $T \geq 3$. The data are fitted to the following four-parameter ansatz :

$$V(R) = V_0 + kR - \frac{e}{R} + g(\frac{1}{R} - [\frac{1}{R}])\,, \qquad (9)$$

where $[\frac{1}{R}]$ is a lattice propagator for the one-gluon exchange (see e.g [20]). Table 5 shows the results for three of our sea-quark masses with estimates of the lattice spacings $a_{q\bar{q}}$ obtained from $k = a^2\sigma$ with $\sqrt{\sigma} = 440$ MeV. These results are preliminary; errors are purely statistical and obtained from a bootstrap analysis with a sample size of 100. The data are shown in Fig. 7. In the future, with substantially improved statistics, we will compare these findings to lattice spacings obtained from the inter-quark force radius à la Sommer[21].

Table 5
Estimates of the lattice spacings obtained from the string tension (preliminary).

| $\kappa_{\rm sea}$ | # of configs | k | $a^{-1}[GeV]$ |
|---|---|---|---|
| 0.156 | 6 | 0.0608(44) | 1.8(2) |
| 0.157 | 29 | 0.0437(18) | 2.1(1) |
| 0.1575 | 12 | 0.0340(18) | 2.4(1) |

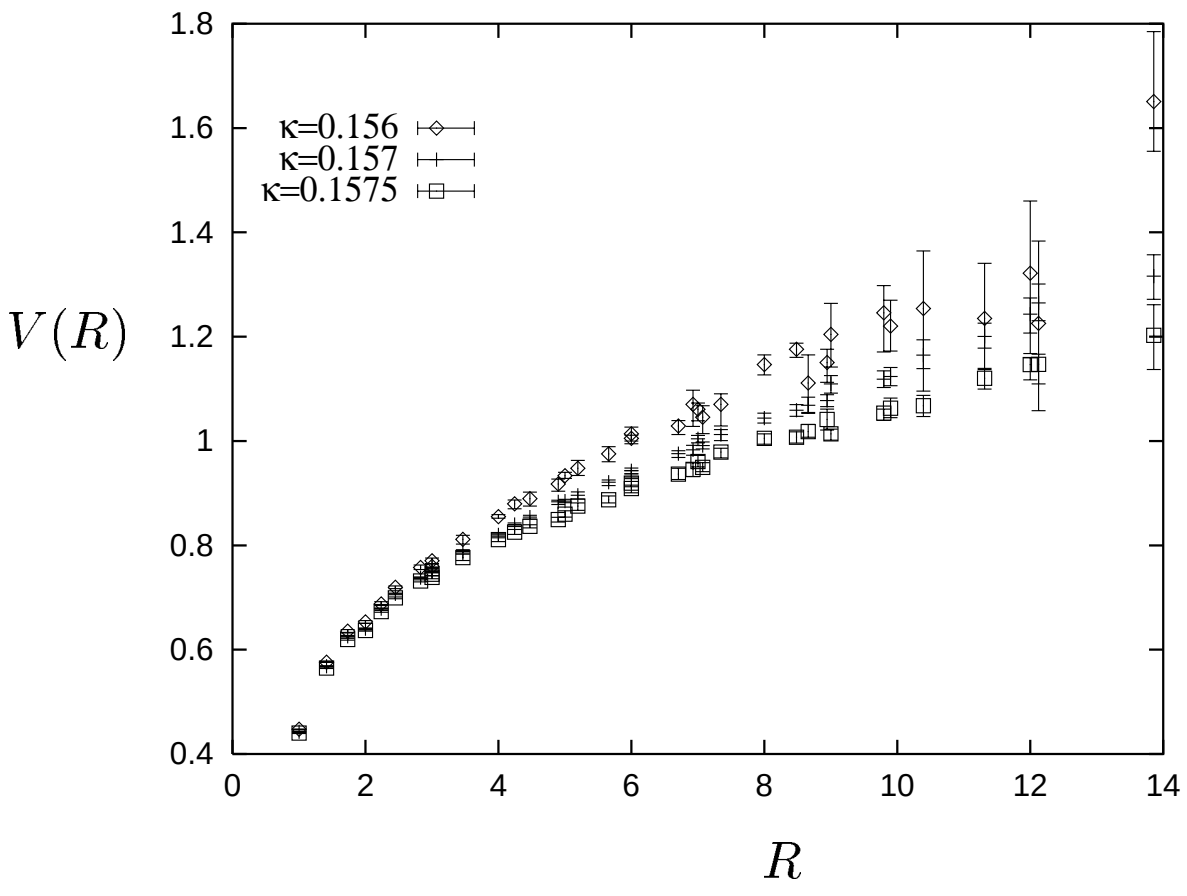

Figure 7. The static potential (in lattice units) for the three $\kappa_{\rm sea}$-values (preliminary).

## 5.3. Light Spectrum

As part of an ongoing study of the light-hadron spectrum we present preliminary results for the lattice spacing obtained from the mass of the rho.

This simulation is being performed with a set of 10 combinations of 4 valence quarks $(\kappa_1, \kappa_2)$ on the three different sea-quark configuration sets. As above, we use decorrelated trajectories separated by more than twice the integrated pion autocorrelation time. We calculate smeared-local and smeared-smeared combinations of correlators, performing 50 smearing-iterations and using 'Gaussian' Wuppertal smearing[19].

So far, we have analysed 4 configurations at $\kappa = 0.157$ so that correlated fits are not yet possible over a large t-range. Using only the smeared-smeared data, we perform linear chiral extrapolations in $m_\pi^2$ to find $\kappa_{\rm crit} = 0.1596$ and also in $m_\rho$ to find a preliminary lattice spacing $a_{m_\rho}^{-1} = 2.48(20) GeV$. These fits are shown in figure 8.

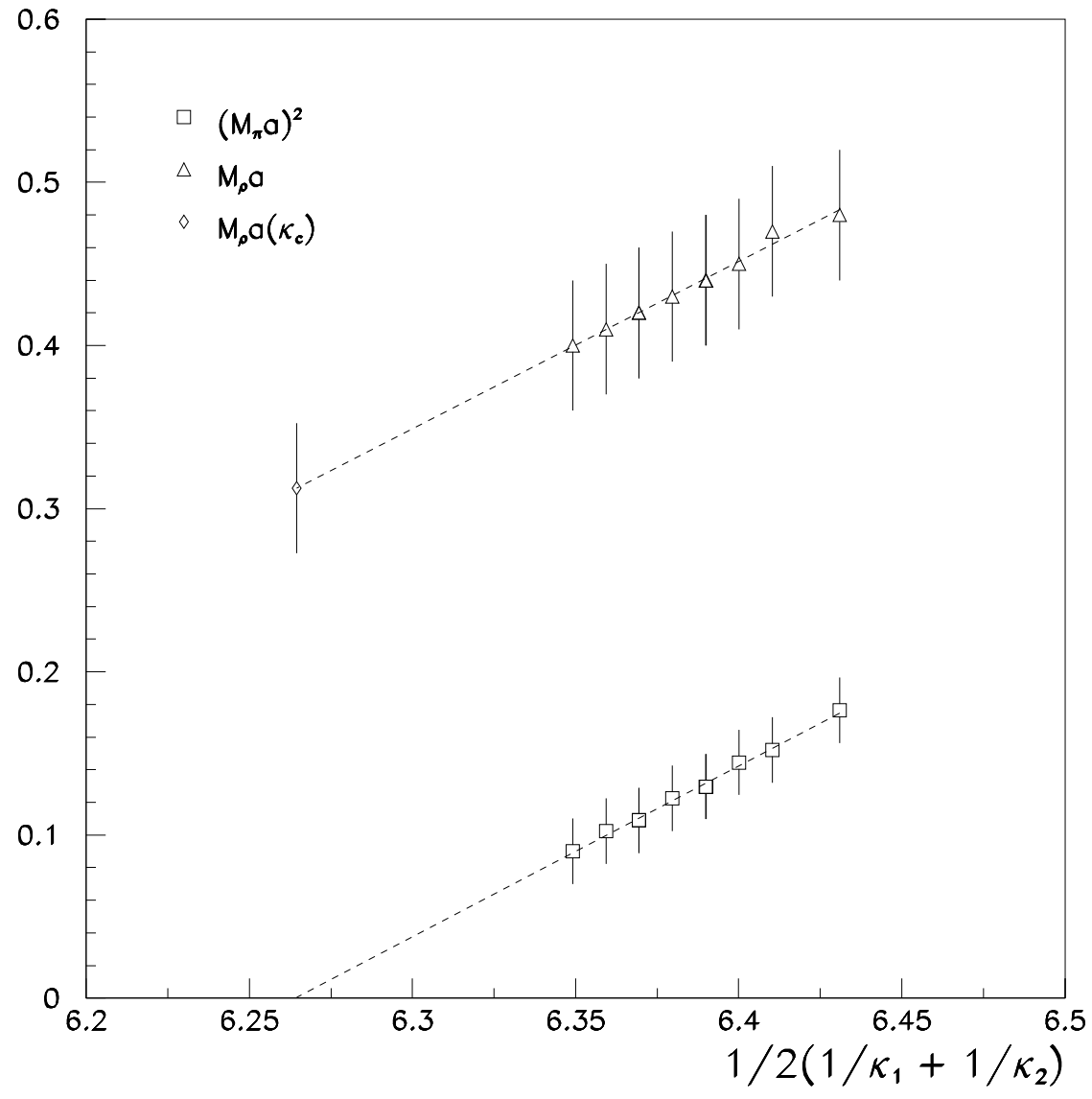

Figure 8. Linear extrapolations of $m_\pi^2$ and $m_\rho$ for a set of 10 combinations of 4 bare quark masses at fixed sea-quark $\kappa_{\rm sea} = 0.157$.

## 6. Conclusions

The main conclusion at this stage of the work is positive: the autocorrelation time in the HMC series is well below 100, enabling us to achieve—in 140 TFlops×hours—a decent statistics on the target sample of 100 independent field configurations.

## Acknowledgements

The project is supported—within the HLRZ—by DESY/Zeuthen and ZAM at KFA/Jülich as well as by IAI/Wuppertal and the Deutsche Forschungsgemeinschaft Research Focus "Dynamical Fermions" (grants Schi 257/1-4 and Schi 257/3-3, Schi 257/5-1, Pe 340/6-1). The work was carried out within the EU network CHRX-CT92-0051. We wish to thank the HLRZ user support groups of DESY/Zeuthen and ZAM of KFA as well as M. Plagge and H. Simma for their assistance.